\documentclass[11pt]{amsart}
\usepackage{geometry}                
\usepackage{graphicx}
\usepackage{amssymb}
\usepackage{epstopdf}
\usepackage{appendix}
\title[Financial correlations at ultra-high frequency]{Financial correlations at ultra-high frequency: theoretical models and empirical estimation}
\author{Iacopo Mastromatteo}
\address{Iacopo Mastromatteo\\ {\it SISSA, Via Beirut 2-4, 34014 Trieste, Italy }}
\author{Matteo Marsili}
\address{Matteo Marsili\\ {\it The Abdus Salam International Centre for Theoretical
Physics, Strada Costiera 11, 34014 Trieste, Italy}}
\author{Patrick Zoi}
\address{Patrick Zoi\\ {\it 
Risk \& Capital Management,
Assicurazioni Generali,
Piazza Duca degli Abruzzi 2,
34132 Trieste, Italy}}
\date{}                                           

\begin{document}

\begin{abstract}
A detailed analysis of correlation between stock returns at high frequency is compared with simple models of random walks. We focus in particular on the dependence of correlations on time scales -- the so-called Epps effect. This provides a characterization of stochastic models of stock price returns which is appropriate at very high frequency. 
\end{abstract}
\maketitle

\section{Introduction}

The study of covariances between stocks is a central problem in finance, both to achieve theoretical understanding of market structure \cite{empirical} and to exploit its relevant applications, such as 
portfolio optimization \cite{EltonGruber}. 
With the availability of financial high frequency data, it has become possible to estimate correlations on very short time scales, down to the frequency of individual transactions. As Epps first observed in 1979, the measured correlations between stock prices decrease as sampling frequency of time series grows \cite{Epps1979}. Since then other studies on data coming from
different stock markets \cite{Bonanno2001} \cite{Zebedee2001} and foreign exchange markets \cite{Lundin1999} \cite{Muthuswamy2001} evidenced the persistence of such phenomenon -- called Epps effects -- across different markets.

Understanding the dependence of financial correlations on time scale  has important practical consequences for portfolio management. For example, for large portfolios, the estimation of risk measures at low frequency (e.g. one day) suffers from instabilities, due to the scarcity of data \cite{kondor}. 
Estimates of financial correlations -- and hence of risk measures -- at high frequency can rely on much richer and longer time series and can potentially detect structural changes more efficiently. 
Relating the structure of correlations at longer time scales to that at shorter time scales, provides means of overcoming the information deficiency causing the instability of risk measures. Interestingly, Borghesi {\em et al.} \cite{Borghesi2007} found that the structure of correlations in groups of very liquid stocks, is largely invariant across time scales ranging from 5 minutes to one day. This suggests that estimates of correlations on long time scales from high frequency correlations is in principle feasible.

Transactions in financial markets play two r\^oles:
in principle (i) they impact returns causing price movements, but in practice (ii)  they also allow prices to be known, fixing the market value of a traded security until the next trade takes place.
Correspondingly, two main contributions to the Epps effect have been considered in the literature so far:
the first relates  the Epps effect to genuine lagged correlations, and it arises from (temporary or permanent) impact of individual trades on the price dynamics. The second relates to the fact that price dynamics is not synchronous across stocks (i.e. transactions take place at different times, in principle, on different stocks).\footnote{The finiteness of the tick-size is also a significant source of Epps
effect; its impact has been investigated in \cite{Munnix2010a} \cite{Munnix2010b}.}

Both lagged correlations and asynchronous sampling contribute to the Epps effect, but the relative weight of these two effects is not always easy to assess (see \cite{Reno2003} and \cite{Large2007}): the first aspect to be considered is the fact that 
trading is not  synchronous so that covariance estimation is intrinsically problematic  at high frequency \cite{Scholes1977}.
Lo and MacKinley proposed a solution to this issue based on a "random censorship" model \cite{Lo1990a}, which was able to explain why simple estimators tend to bias correlations towards
zero at high frequency (more recent works following this line are \cite{Toth2009a} \cite{Toth2009b} \cite{Barndorff-Nielsen2009} and \cite{Zhang2010}). The second factor contributing to the Epps effect is the presence of genuinely lagged correlations (lead-lag effect) \cite{Lo1990b} \cite{Kullmann2002} \cite{Toth2006},
which should contain informations about the dynamical structure of the market.

This paper addresses the issue of disentangling these effects at very high frequency. 
We adopt an approach similar to \cite{Lo1990a}, and use a previous tick estimator (see \cite{Dacorogna2001} for an analysis of interpolation-based estimators) to check the impact of asynchronous trading on
correlations, without any specific choice for their genuine structure; alternative choices to deal with asynchrony are indeed available (namely \cite{Malliavin2002} and \cite{Hayashi2005}). The
performance of some popular estimators has been investigated in \cite{Griffin2011}.

We discuss a minimal model of price dynamics, which describes an infinitely 
liquid market: a transaction in this scenario has the only effect of revealing the asset price at a given instant of time, but sampling has no impact on prices. 
We find that also in this oversimplified scenario transactions can strongly affect correlations; in particular the  Epps effect is always dominated by the asynchronous sampling at very high frequency. We show that it is possible to infer the genuine correlation structure of the market if one supposes
inter-trade times to be exponentially distributed; in particular we can analytically disentangle the contribution to the Epps effect due to asynchronous trading to the one due
to a genuine lag.

We apply the model to data of NYSE, finding that some features of the time series of returns at very high frequency can successfully be reproduced. In particular, assuming a process of asynchronous sampling of correlated random walks, we can estimate the underlying correlation function. The heterogeneity of sampling frequency in the bare data implies some predictability of less active stocks from the knowledge of more active ones. But once the effect of asynchronous sampling is removed, we find no causal structure in lagged cross-correlations. Still, cross-correlations are significantly non-zero over time lags of the order of ten seconds, whereas auto-correlations decay on the scale of one or two seconds. This provides evidence of an information contagion process across stocks, at ultra-high frequency.

The rest of the paper is organized as follows: we first discuss the origin of the Epps effect in simple theoretical models with synchronous (Section \ref{SyncSection})  or asynchronous sampling (Section \ref{AsyncSection}). Section \ref{FilterSection} discusses how to reconstruct the underlying correlations from asynchronously sampled data, in theoretical models. Section \ref{EmpirSection} applies these insights to empirical tick-by-tick data of NYSE. We summarize and discuss our results in Section \ref{FinalSection}. Technical derivations and proofs are relegated to the appendix for the sake of readability.

\section{The origin of Epps effect: simple theoretical models}
\label{SyncSection}

Consider a multivariate time series with stationary increments $dX^i_t$, where $t$ is a continuous time parameter and $i=1,\dots,n$. The series will represent in the following
the infinitesimal increment of the log-price of asset $i$ at time $t$; let the finite variation of log-price after a time $\Delta t$ be given by\footnote{All the following considerations can be easily generalized to the discrete time case. We choose for simplicity to present them in continuous time. Notice that $c^{ij}_\tau$ is to be interpreted as a distribution (e.g. it may contain terms proportional to $\delta(\tau)$).} :
\[
X^i_{\Delta t} = \int_0^{\Delta t} dX^i_t \;,
\]
and say that the infinitesimal, lagged correlations are given by:
\[
c^{ij}_{t-t^\prime}\,dt\,dt^\prime = \langle dX^i_t dX^j_{t^\prime} \rangle
\]
while the spectrum $S^{ij}_\omega$ is defined as
\[
S^{ij}_\omega = \int_{-\infty}^{+\infty} d\tau \, c^{ij}_\tau \, e^{i \omega \tau}.
\]

We will be interested in characterizing the dependence  of the finite, equal time correlation $C^{ij}_{\Delta t} = \langle X^i_{\Delta t} X^j_{\Delta t} \rangle$ on the time scale $\Delta t$.
Its behavior can be extracted from the knowledge of the series $dX^i_t$, which can be related to $C^{ij}_{\Delta t}$ as:
\begin{eqnarray}
\label{Covariance}
C^{ij}_{\Delta t} &=& \int_0^{\Delta t} \int_0^{\Delta t} dt \, dt^{\prime} \, c^{ij}_{t-t^\prime} \\
&=& \frac{1}{2 \pi} \int_{-\infty}^{+\infty} d \omega \, \frac{S^{ij}_\omega}{\omega^2} \left( e^{- i \omega \Delta t} -1 \right) \left( e^{ i \omega \Delta t} -1 \right) \nonumber
\end{eqnarray}
While for a purely Brownian motion, the scaling of $C^{ij}_{\Delta t}$ is linear in $\Delta t$, in the general case we will quantify deviations from linearity of $C^{ij}_{\Delta t}$ by considering the quantity:
\begin{equation}
\rho^{ij}_{\Delta t} = \frac{C^{ij}_{\Delta t}}{\sqrt{C^{ii}_{\Delta t} C^{jj}_{\Delta t} }} \;,
\end{equation}
that is the Pearson correlation coefficient, built by normalizing the covariance to the variances. We will say that {\em the Epps effect is absent whenever $\rho^{ij}_{\Delta t}$ is independent of $\Delta t$, and it is present otherwise.}\\

It is interesting to remark some general features of $\rho^{ij}_{\Delta t}$: first, the positivity of the eigenvalues of the covariance matrix $C^{ij}_{\Delta t}$ ensures that $|\rho^{ij}_{\Delta t}| \leq 1$. The finiteness of
the limit $\Delta t \rightarrow 0$ of $\rho^{ij}_{\Delta t} $ can then be checked from the continuity of the coefficients: if both auto- and cross-correlations are infinite at the origin, then $\rho^{ij}_{\Delta t}$ is finite; the same holds
if auto- and cross-correlations are both finite at the origin. The case with infinite auto-correlations and finite cross-correlations gives instead $\rho^{ij}_{\Delta t} \rightarrow 0$: whenever the time needed by the system to
auto-correlate is much smaller than the time needed to cross-correlate, then the equal time correlation coefficient goes to 0.
In the opposite limit $\Delta t \rightarrow \infty$, if $\int_{-\infty}^{+\infty} d\tau \, c^{ij}_\tau = \mathcal C ^{ij}$ is finite, one can also see that:
\[
\rho^{ij}_\infty = \frac{ \mathcal C^{ij}}{\sqrt{ \mathcal C^{ii} \mathcal C^{jj}}}
\]
The behavior of $\rho^{ij}_{\Delta t}$ during the transient is also interesting, as it contains non-trivial informations about the time needed by the system to correlate the dynamics. The origin of the Epps effect is best illustrated by discussing few simple examples.

\subsection{Example (Correlated Brownian motions):}
Let's consider the case of a bivariate process of the kind:
\begin{eqnarray}
dX^1_t&=&\sqrt{c} \, d\eta^0_t + \sqrt{1-c} \, d\eta^1_t \nonumber \\
dX^2_t&=&\sqrt{c} \, d\eta^0_t + \sqrt{1-c} \, d\eta^2_t \nonumber
\end{eqnarray}
where the $d\eta_t^i$ are white noises, so that $\langle d\eta_t^i d\eta^j_{t^\prime} \rangle =$ \\ $ \delta^{ij} \delta_{t-t^\prime} \,dt\, dt^\prime$. Then this is the only case in which linearity strictly holds both for variance and covariance:
\begin{eqnarray}
C^{12}_{\Delta t} &=& c \, \Delta t \nonumber \\ 
C^{ii}_{\Delta t} &=& \phantom{c \,} \Delta t  \nonumber
\end{eqnarray}
so that:
\[
\rho^{12}_{\Delta t} = c  \;, 
\]
independent of $\Delta t$, and there is no Epps effect.

\subsection{Example (Lagged series):}
\label{ExampleLagged}
Let's now consider the lagged version of the previous process:
\begin{eqnarray}
dX^1_t&=&\sqrt{c} \, d\eta^0_{t\phantom{+\tau}} + \sqrt{1-c} \, d\eta^1_t \nonumber \\
dX^2_t&=&\sqrt{c} \, d\eta^0_{t+\tau} + \sqrt{1-c} \, d\eta^2_t \nonumber
\end{eqnarray}
In this case:
\begin{eqnarray}
c_{t-t^\prime}^{ii} &=& \phantom{c}\, \delta_{t-t^\prime} \nonumber \\
c_{t-t^\prime}^{12} &=& c \, \delta_{t-t^\prime -\tau} \nonumber
\end{eqnarray}
and it is easy to see (appendix \ref{AppendixA}) that $\rho$ results in this case:
\[
\rho^{12}_{\Delta t} = c \left(  1  - \frac{ \tau}{\Delta t} \right) \theta(\Delta t - \tau) \;, 
\]
where $\theta (t)$ is the step function, so the presence of an Epps effect is evident.

\subsection{Example (Different widths):}
We can now consider another bivariate process, whose lagged correlations are:
\begin{eqnarray}
c_{t-t^\prime}^{12} &=& c \, \left( \frac{1}{2\, \xi_l} e^{- |t - t'| /\xi_l} \right) \nonumber \\
c_{t-t^\prime}^{ii} &=& \phantom{c \, (} \frac{1}{2 \,\xi_s} e^{- |t - t'| /\xi_s} \,, \nonumber
\end{eqnarray}
with the conditions $\xi_l \geq \xi_s$ and $c \leq 1$ ensuring $|\rho^{12}_{\Delta t}| \leq 1$. In this case one has:
\[
\rho^{12}_{\Delta t} = c \left[ \frac{\Delta t + \xi_l \left( e^{-\Delta t/\xi_l} - 1\right)}{\Delta t + \xi_s \left( e^{-\Delta t/\xi_s} - 1\right)} \right] \nonumber
\]
Such quantity is a constant only for $\xi_s = \xi_l$, while in the general case it is a function which grows from $\rho^{12}_0 = c \, \xi_s / \xi_l$ to an asymptotic value $\rho^{12}_\infty$, as represented
in the blue lines of figure \ref{Fig3}. The case $\xi_s \rightarrow 0$ is also interesting, as the variance becomes linear, while $\rho^{ij}_{\Delta t}$ is given by:
\[
\rho^{12}_{\Delta t} = c \left[ 1 + \frac{\xi_l}{\Delta t} \left( e^{-\Delta t / \xi_l} - 1\right) \right] \nonumber
\]

The above examples show that an Epps effect is present if the covariance of a process grows with $\Delta t$ at a rate smaller than the variance, or equivalently the infinitesimal, lagged cross-correlations $c^{ij}_\tau$ are not proportional to auto-correlations $c^{ii}_\tau$. We will see in section \ref{EmpirSection} that financial time series
show at high frequency a correlation structure which is reminiscent of the one of these examples; in particular such structure is well fitted by a model where the dynamics of correlations
is described by a lag parameter $\tau$ and a width parameter $\xi$, and where variances grow faster with respect to covariances. In \cite{Bouchaud2005} this approach
is also used to describe the dynamics related with the time evolution of the correlation matrix.

\section{Asynchronous sampling of correlated random walks}
\label{AsyncSection}

While studying a multivariate time series at very high frequency (say, tick-by-tick financial data), it is unlikely that all transactions happen simultaneously; additionally some time bin may contain no
data point at all, as no transaction took place. This fact may cause problems in the estimation of volatilities and correlations  \cite{Scholes1977}, especially in the extreme case in which one tries to evaluate
such quantities at time scales of the order of the inter-trade time. A possible approach to deal with this issue is the creation of a synchronous series \cite{Lo1990a} out of the asynchronous one by means of some
prescription, such as linear interpolation or previous-tick interpolation \cite{Dacorogna2001}.
We adopt this latter, simpler estimator to study the impact of asynchrony on measured correlations, as it allows an easy analytical treatment of such quantities without requiring any assumption on their
genuine nature (in particular we will focus on models containing lagged and short ranged correlations).

\begin{figure}
\begin{center}
\includegraphics[width=3in]{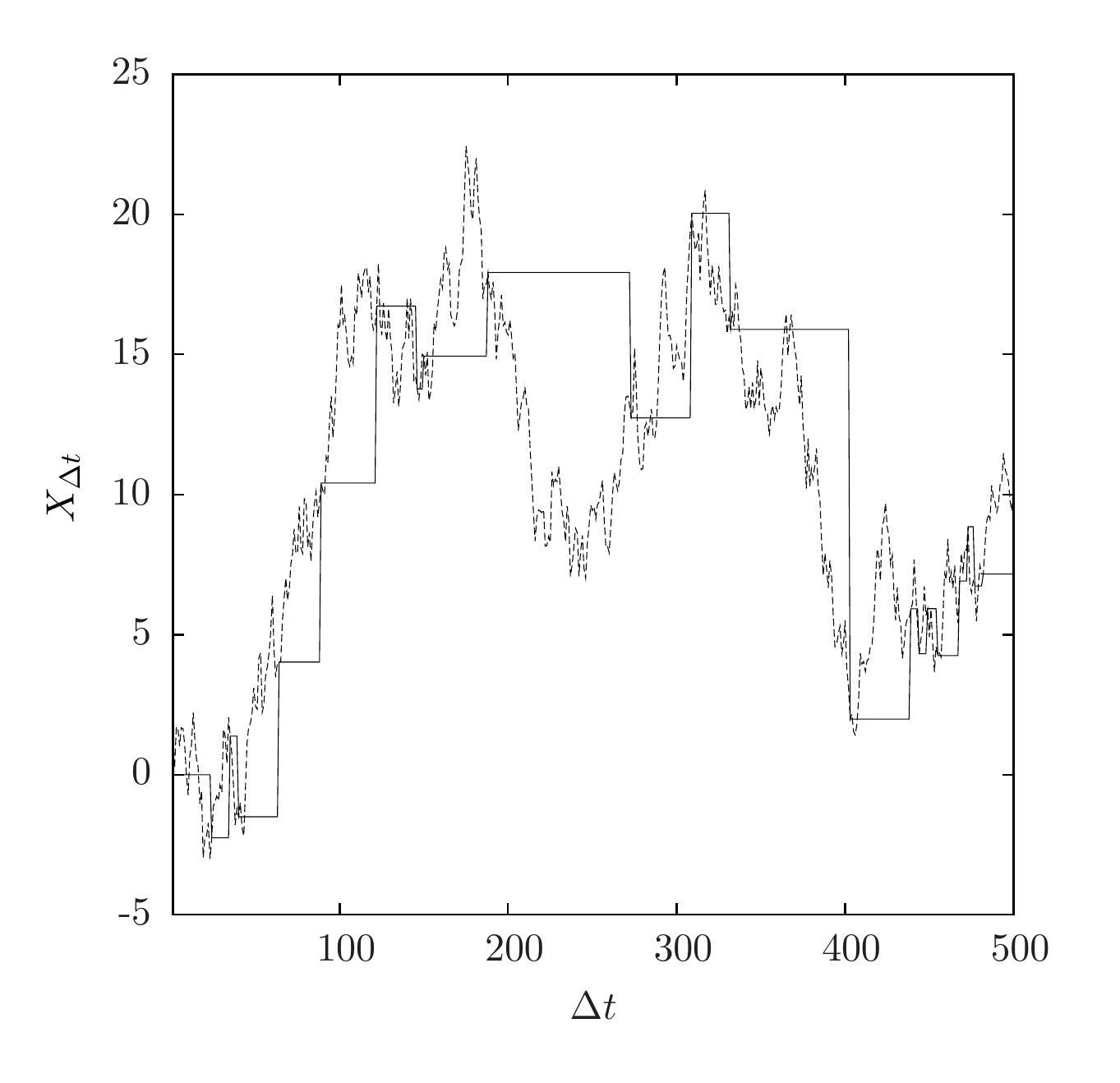}
\caption{We plot here a realization of a synchronous process $X_{\Delta t}$ (dashed line), and a randomly sampled version of the same realization (full line), obtained with a sampling rate $\lambda = 0.05$.}
\label{Fig1}
\end{center}
\end{figure}

Consider an underlying synchronous process $dX^i_t$ defined as in section \ref{SyncSection}, and $n$ subsets of points $U^i = \{t^i_k\}_{k \in \mathbb{Z}}$ randomly drawn on the real line. Let the probability
of drawing a point between $t$ and $t+dt$ for subset $U^i$ be given by $\lambda_i \, dt$. In this way for each subset $U^i$ the number of points drawn in an interval $[t_1,t_2]$ is a Poissonian random variable
of mean $\lambda_i \, (t_2 - t_1)$. The corresponding waiting time distribution is exponential, and is given by $p_i(t) = \lambda_i e^{-\lambda_i t}$.
Given a set of $U^i$ and a realization of the underlying synchronous process $X^i_t$, one can define an asynchronous process:
\[
\tilde X^i_{\Delta t} = \int_{t_1}^{t_2} d X^i_t
\]
where $t_1 = \max \{ t^i_k \in U^i |\,  t^i_k < 0 \}$ and  $t_2 = \max \{ t^i_k \in U^i |\,  t^i_k < \Delta t \}$. This time series is a piecewise constant function, with discrete jumps at the points $\Delta t = t^i_k$ , as shown in
figure \ref{Fig1}); notice that this construction implements the \emph{previous tick estimator} prescription (PTE) to deal with missing data.
Covariance can be defined in this case as:
\[
\tilde C^{ij}_{\Delta t} = E \left[ \langle \tilde X^i_{\Delta t} \tilde X^j_{\Delta t} \rangle \right]
\]
where $E[ \cdot ]$ denotes expectation value with respect to the sampling process. Then one can generalize the Epps effect, defining as in the previous case the function $\tilde \rho^{ij}_{\Delta t} $:
if such function depends on $\Delta t$, (generalized) Epps effect is present, otherwise it is absent. \\ \\
We will now show three properties which allow to extract information about the asynchronous process $\tilde X^i_{\Delta t}$ given the spectrum of the synchronous process $X^i_{\Delta t}$.
The proof of these results is given in appendix \ref{AppendixB}.
\subsubsection*{{\bf P1:} Covariance of asynchronous processes}
Given an asynchronous time series $\tilde X_{\Delta t}^i$ defined using a synchronous time series of spectrum $S^{ij}_\omega$ and waiting time distributions $p^i(t) = \theta(t) \lambda_i \, 
e^{-\lambda_i t}$, for $i \neq j$ it holds:
\begin{equation}
\tilde C^{ij}_{\Delta t} = \frac{1}{2 \pi} \int_{-\infty}^{+\infty} d \omega \, \frac{S^{ij}_\omega }{\omega^2} \, \left[  \frac{\lambda_i \lambda_j}{(\lambda_i + i \omega)(\lambda_j - i \omega)} \right] 
\left( e^{- i \omega \Delta t} -1 \right) \left( e^{ i \omega \Delta t} -1 \right)
\label{AsyncCovariance}
\end{equation}

Equivalently, covariance in the asynchronous case can be computed by correcting the synchronous spectrum with the substitution:
\begin{equation}
\tilde S^{ij}_{\omega} = S^{ij}_{\omega}  \frac{\lambda_i \lambda_j}{(\lambda_i + i \omega)(\lambda_j - i \omega)}  \label{SpectrumSubstitution}
\end{equation} \\
In real space, such substitution is equivalent to the convolution:
\begin{equation}
\tilde c_{t-t^\prime}^{ij}=  \frac{\lambda_i \lambda_j}{(\lambda_i + \lambda_j)} \Bigg[
\int_{-\infty}^{t'} d\tau \;  c_{t-\tau}^{ij} e^{- \lambda_j (t'-\tau)}
+ \int_{t'}^{+\infty} d\tau \;  c_{t-\tau}^{ij} e^{- \lambda_i (\tau - t^\prime)} \Bigg] .
\label{Convolution}
\end{equation}
\subsubsection*{{\bf P2:} Variance of asynchronous processes}
Consider the asynchronous time series $\tilde X_{\Delta t}$, defined using a synchronous time series of spectrum $S_\omega$ and a waiting time distribution $p(t) = \theta(t) \lambda \, e^{-\lambda t}$.
Then it holds:
\begin{eqnarray}
\tilde C_{\Delta t} &=& \frac{1}{2 \pi} \int_{-\infty}^{+\infty} d \omega \, \frac{S_\omega}{\omega^2} \left( e^{- i \omega \Delta t} -1 \right) \left( e^{ i \omega \Delta t} -1 \right) + \nonumber \\
&+& \frac{2}{\lambda^2} \left[  \frac{1}{2 \pi} \int_{-\infty}^{+\infty} d \omega \, \frac{S_\omega}{1+ \omega^2/\lambda^2}  \left( e^{-i \omega \Delta t} - e^{-\lambda \Delta t} \right) \right] \nonumber
\end{eqnarray}
Equivalently, to compute the variance in the asynchronous case it is necessary to add to the synchronous value a correction, so that one gets:
\begin{equation}
\label{VarianceSubstitution}
\tilde C_{\Delta t} = C_{\Delta t} + \frac{2}{\lambda^2}\left( \bar c_{\Delta t} - e^{-\lambda \Delta t} \bar c_0 \right) \,
\end{equation}
where $\bar c_\tau$ is the Fourier anti-transform of the damped spectrum $ \frac{S_\omega}{1+\omega^2 / \lambda^2}$.
\subsubsection*{{\bf P3:} Case of linear variance} If an asynchronous time series $\tilde X_{\Delta t}$ is defined using a synchronous series of variance $C_{\Delta t}$ linear in $\Delta t$
 (corresponding to a constant spectrum $S_\omega$)
and waiting time distribution $p(t) = \theta(t) \lambda \, e^{-\lambda t}$, then the asynchronous value of its variance corresponds to the synchronous one.\\ \\
The property P1 shows what is the effect of the random sampling on the measured covariance: if $\lambda=\lambda_i=\lambda_j$ substitution (\ref{SpectrumSubstitution}) is a low-pass filter (a
Lorentzian) with a cutoff scale set by $\lambda$, which suppresses signal at frequencies bigger than the sampling scale. In the case $\lambda_i \neq \lambda_j$ an effect of spurious 
causality\footnote{We employ the term "causality" in a loose sense, using the expression "returns of stock $i$ cause returns of stock $j$" to signify that $c^{ij}_\tau > c^{ij}_{-\tau}$, that is, an asymmetry
is measured in the lagged correlation of two stocks} is also induced: kernel (\ref{SpectrumSubstitution})
has in general a complex phase, which generates an asymmetry between $\tilde c^{ij}_\tau$ and $\tilde c^{ij}_{-\tau}$, as pointed out in \cite{Lo1990a}. The direction of such asymmetry is such that the
more frequently sampled series appears to influence the less sampled
one: this merely reflects the fact that one can use the information contained in the more sampled series to successfully forecast the less sampled one. The property P2 allows to calculate in general 
the asynchronous value of the variance, and in particular for the simple case of a very narrow correlation coefficient $c^{ii}_\tau$ P3 implies that variance doesn't necessarily decrease as $\lambda_i$ 
gets smaller, while covariance always gets suppressed; this is why, generally speaking, asynchronous sampling tends to enhance Epps effect.
Notice that, while P1 directly relates $c^{ij}_\tau$ with $\tilde c^{ij}_\tau$ (and $S^{ij}_\omega$ with $\tilde S^{ij}_\omega$), P2 just connects $C^{ii}_{\Delta t}$ with $\tilde C^{ii}_{\Delta t}$:
the asynchronous value of the auto-correlation function $\tilde c^{ii}_\tau$ has to be indirectly obtained from $\tilde C^{ii}_{\Delta t}$. For $\tau \neq 0$, this can be done by observing that:
\begin{equation}
\label{VarSpectrum}
 \frac{d^2}{d\Delta t^2} \tilde  C^{ii}_{\Delta t} \Bigg|_{\Delta t = \tau} = \tilde  c_\tau^{ii} + \tilde  c_{-\tau}^{ii}
 = 2 \tilde c^{ii}_\tau = \frac{1}{\pi} \int d\omega \, \tilde S^{ii}_\omega \, e^{-i \omega \tau}
\end{equation}
For $\tau = 0$ instead the auto-correlation $\tilde  c^{ii}_\tau$ may contain a $\delta_\tau$, which can be deduced from the behavior of $\tilde C^{ii}_{\Delta t}$ for $\Delta t \rightarrow 0$. Specifically, 
if $\tilde C^{ii}_{\Delta t} \sim \Delta t$, then the auto-correlation is divergent in $\tau = 0$, which signals the presence of a term $\delta_\tau$. Conversely,  if $\tilde C^{ii}_{\Delta t} \sim \Delta t^2$ or if $\tilde C^{ii}_{\Delta t}$ vanishes faster than $\Delta t^2$, then $\tilde c_\tau^{ii}$ is regular in 0.

These results allow us to generalize the analysis of the examples in section \ref{SyncSection} to the asynchronous case.
\subsection{Example (Correlated Brownian motions):}
In this simple case the synchronous value of the correlation coefficient is given by:
\[
c_\tau^{12} =  c \, \delta_\tau \,
\]
If now we suppose the rates of the sampling processes to be all equal to $\lambda$, equation (\ref{Convolution}) can be 
used to calculate the asynchronous value of covariance, while the variance inherits linearity from the synchronous case.
The result reads (see appendix \ref{AppendixA}):
\[
\tilde \rho^{12}_{\Delta t} = c \, \left( 1 + \frac{1}{\lambda \Delta t} \left( e^{- \lambda \Delta t } - 1\right) \right) \;,
\]
which is plotted in figure \ref{Fig3} (black line). In this case we have a spurious (induced by the sampling) Epps effect, as the original time series did not show any Epps effect.
\subsection{Example (Lagged series):}
\begin{figure}
\begin{center}
\includegraphics[width=3in]{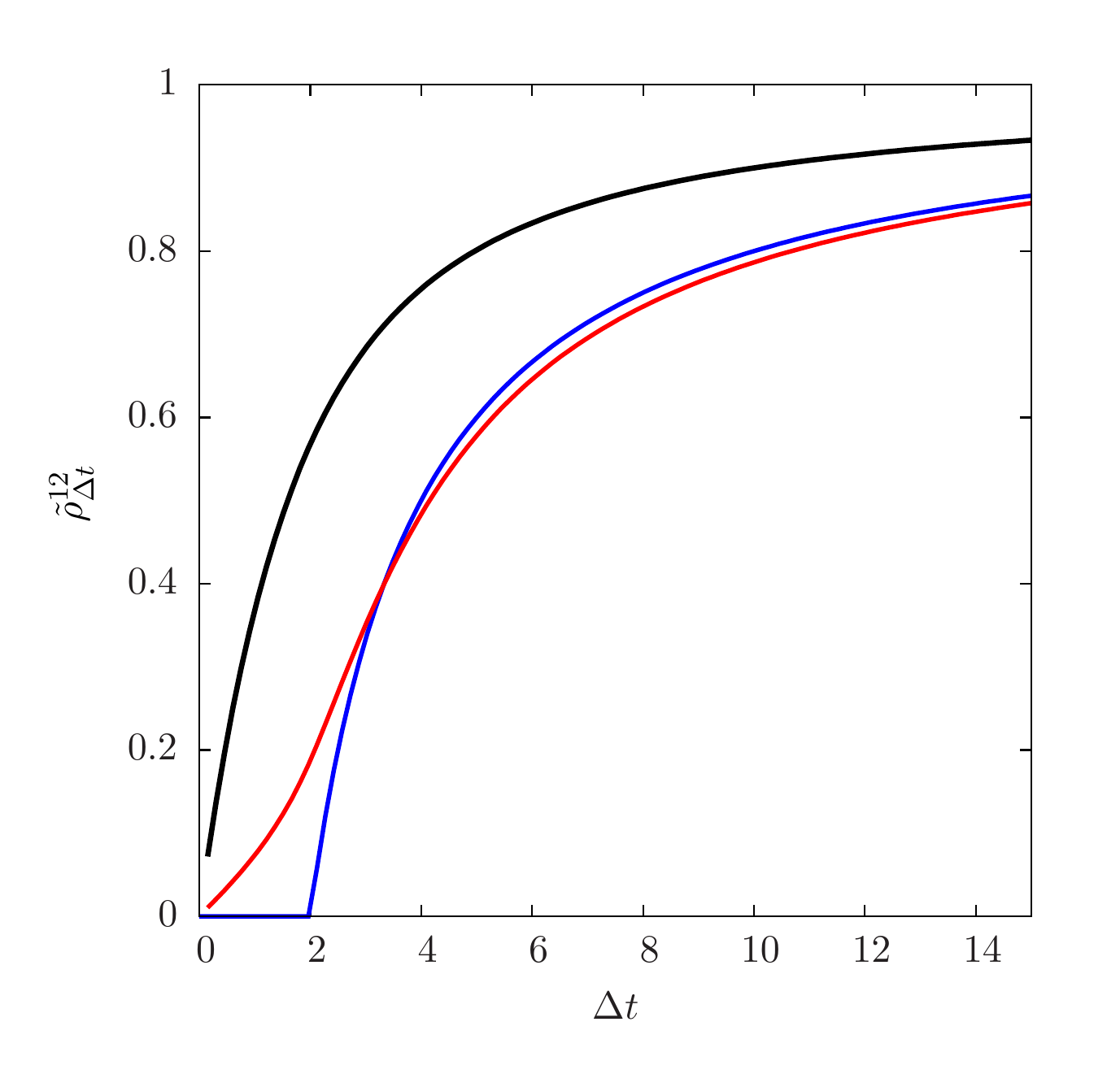} 
\caption{Equal time correlation coefficient for two lagged processes, both for the case of synchronous and asynchronous sampling. The lag parameter $\tau$ and the sampling rate $\lambda$
are set to $\tau=0,\lambda=1$ (black line),  $\tau=2,\lambda=\infty$ (blue), $\tau=2,\lambda=1$ (red line).}
\label{Fig2}       
 \end{center}
\end{figure}

Now we turn to the synchronous process:
\begin{eqnarray}
dX^1_t&=&\sqrt{c} \, d\eta^0_{t\phantom{+\tau}} + \sqrt{1-c} \, d\eta^1_t \nonumber \\
dX^2_t&=&\sqrt{c} \, d\eta^0_{t+\tau} + \sqrt{1-c} \, d\eta^2_t \nonumber
\end{eqnarray}
and consider again sampling rates $\lambda_1 = \lambda_2 = \lambda$. Then,  using the above properties, one can find that (see appendix \ref{AppendixA}):
\[ \tilde \rho^{ij}_{\Delta t} =\left\{
\begin{array}{lcc} \displaystyle{\frac{c}{2\lambda \Delta t} e^{-\lambda (\Delta t+\tau)}\left(1-e^{\lambda \Delta t}\right)^2  }& \rm{if } &\Delta t < \tau\\
\phantom{,}&& \\
\displaystyle{\frac{c}{\lambda \Delta t} \left[e^{-\lambda \Delta t} \cosh(\lambda \tau) - e^{-\lambda \tau}\right] +}  && \\
\\
\displaystyle{c\left(1 - \frac{\tau}{\Delta t}\right)}&  \rm{if } & \Delta t > \tau \end{array} \right.
\]
and check that Epps effect is enhanced by the effect of the sampling (covariance grows even slower than in the synchronous case), so that genuine and spurious effects
superimpose as shown in figure \ref{Fig2}.
\subsection{Example (Different widths):} 
\begin{figure}
 \begin{center}
  \includegraphics[width=\linewidth]{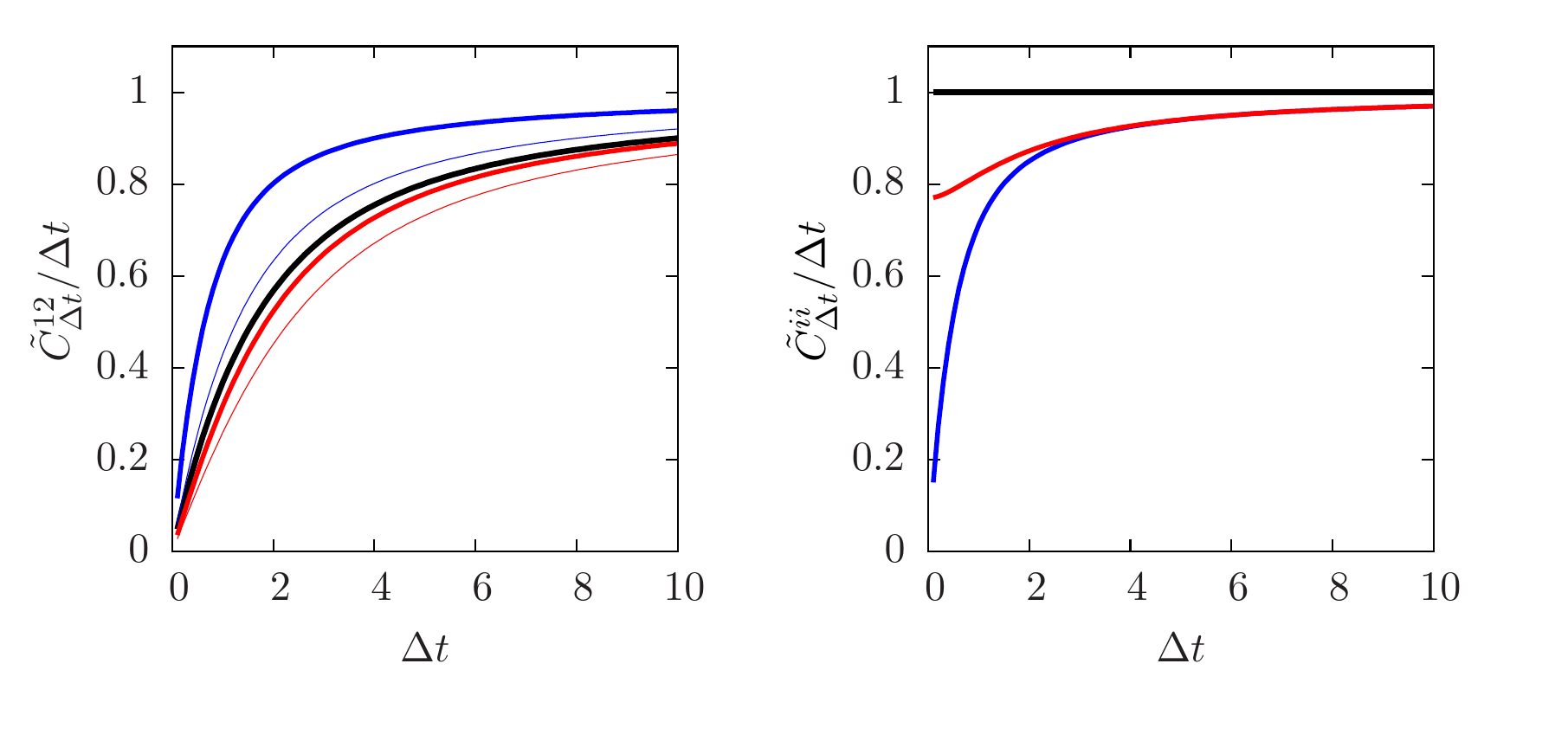}
\caption{Normalized covariance and variance for two processes displaying exponential decay both of cross-correlation and of auto-correlation, where the decay constant are respectively
$\xi_l$ and $\xi_s$, with $\xi_s < \xi_l$. Both the case of synchronous and asynchronous sampling (of common rate $\lambda$) are represented. On the right, the variance is plotted
in the cases $\lambda=\infty$, $\xi_s = 0.3$ (blue line), $\lambda=1$, $\xi_s =0$ (black line) and $\lambda=1$, $\xi_s =0.3$ (red line). On the left, the covariance is represented in the cases
 $\lambda=\infty$, $\xi_l = 0.4$ (thick blue line), $\lambda=\infty$, $\xi_l = 0.8$ (narrow blue line), $\lambda=1$, $\xi_l = 0.4$ (thick red line), $\lambda=1$, $\xi_l = 0.8$ (narrow red line)
 and  $\lambda=1$, $\xi_l = 0$ (black line).}
\label{Fig3}
\end{center}
\end{figure}

Also in this case the genuine Epps effect is enhanced by the asynchronous sampling; indeed in this case both variance and covariance
are influenced by the sampling and produce a spurious effect.
It is possible to calculate the coefficient $\tilde C^{12}_{\Delta t}$ assuming sampling rates $\lambda_1$ and $\lambda_2$. Its value reads:
\[
\tilde C^{12}_{\Delta t} =   c \,  \Bigg[ \Delta t+ \bigg( \frac{\lambda_1 \lambda_2 \xi_l^3}{2 u_1 v_2} (e^{- \Delta t / \xi_l} - 1)  
- \frac{\lambda_2}{\lambda_1(\lambda_1+\lambda_2)u_1 v_1} 
(e^{- \lambda_1 \Delta t}-1) \bigg)  
+ \bigg( \lambda_1 \leftrightarrow \lambda_2 \bigg) \Bigg] 
\]
where the coefficients $u_i$ and $v_i$ are defined in appendix \ref{AppendixA}. The variance is given by:
\[
\tilde C^{ii}_{\Delta t} = \Delta t + \xi_s \left( \frac{\lambda_i^2 \xi_s^2 (e^{-\Delta t/\xi_s} -1) -(e^{-\lambda_i \Delta t}-1)}{\lambda_i^2 \xi_s^2-1} \right)
\]
Notice that the sampling induces a singular
auto-correlation function for the variance: while the synchronous value of $c^{ii}_\tau$ is regular in the origin, one can check that it becomes singular in zero as an effect of the sampling.
In particular, using equation (\ref{VarSpectrum}), one finds that the asynchronous auto-correlation is given by:
\[
\tilde c^{ii}_\tau = \frac{1}{1+\lambda_i \xi_s} \, \delta_\tau + \frac{\xi_s \lambda_i^2}{2(\lambda_i^2 \xi_s^2 - 1)} \left( e^{-|\tau| / \xi} - e^{-\lambda_i |\tau|} \right) \;,
\]
and it is easy to see that the regular part goes to zero for small values of $\tau$, a feature which is also present in empirical data. \\
This example shows how the Epps effect can be induced both from variance and covariance (as in this case neither of those quantities is linear), and that the sampling may
additionally give a spurious contribution to the Epps effect (as the functional dependence of $\tilde C^{ij}_{\Delta t}$ and $\tilde C^{ii}_{\Delta t}$ changes due to the sampling).  In
 figure  \ref{Fig3} some typical curves for normalized variance and covariance are presented for this model.

\section{Filtering of asynchronous time series}
\label{FilterSection}

An interesting application of property (\ref{Convolution}) concerns data filtering of asynchronous time series: as it is possible to quantify how a synchronous time series is influenced by an exponential random sampling, it is also possible to discount its damping effect on the high frequency region of the cross-correlation spectrum. As the random sampling induces a convolution with a known kernel, the reconstruction of the genuine cross correlation structure requires a deconvolution.
In particular, given a measured asynchronous time series $\tilde X^i_t$, the deconvolution procedure can be carried on following these lines:
\begin{enumerate}
\item{Calculate the measured spectrum $\tilde S^{ij}_\omega$ of the time series from raw data $\tilde X^i_t$}
\item{Compute the sampling rate $\lambda_i$ for each process;}
\item{Estimate the genuine spectrum $\hat S_\omega^{ij}$\footnote{Notice that the corrected spectrum has the right properties to construct consistent correlation matrices;
in particular it is Hermitian and it satisfies $\hat S^{ji}_\omega = \hat S^{ij}_{-\omega}$.} by inverting (\ref{SpectrumSubstitution}):
\[
\hat S_\omega^{ij} = \tilde S_\omega^{ij}  \left(  \frac{\lambda_i \lambda_j}{(\lambda_i + i \omega)(\lambda_j - i \omega)} \right)^{-1} = \tilde S_\omega^{ij} K^{ij\,-1}_\omega
\]
}
\item{Write cross-correlations $\hat c_{t-t^\prime}^{ij}$ using equation (\ref{Covariance})}
\end{enumerate}
This deconvolution procedure, known as \emph{inverse filtering}, should in principle allow to compute the genuine signal with infinite accuracy; in practice, dealing with time series of finite length
and in which time is discretized, some effects have to be taken into account. Moreover, while effects of discreteness and finite size are easy to quantify (appendix \ref{AppendixC}), a more careful
treatment of noise is needed: as the inverse deconvolution amplifies the high frequency region of the spectrum with a term proportional to $\omega^2$, the noise that typically dominates that region
affects crucially the accuracy of the reconstructed signal. A possible solution is to set a cutoff to the maximum frequency used to deconvolve the spectrum, choosing for example a deconvolution kernel
of the kind:
\[
\hat S^{ij}_\omega = \tilde S^{ij}_\omega K^{ij\,-1}_\omega \left( \frac{|K^{ij}_\omega|^2}{|K^{ij}_\omega|^2 +{\rm SNR}_\omega^{-1\,ij}} \right)
\]
where $\rm{SNR}^{ij}_\omega$ is the expected signal-to-noise ratio of the genuine signal. This leads to what is called a \emph{Wiener filter} \cite{Wiener1949}.
\\ \\

\section{Empirical analysis on NYSE data}
\label{EmpirSection}

An empirical analysis has been carried on using tick-by-tick data from the New York Stock Exchange (NYSE) collected during the period going from 02.01.2003 to \\
12.31.2003. We studied daily
time series ($T=20000$ seconds) of the 100 most traded stocks, excluding for each day the first 45 and the last 21 minutes of trades, and then averaged over a set of $M=248$ days to obtain the
spectra $\tilde S_\omega^{ij}$. $\tilde X^i_{\Delta t}$ was computed from the observed values of $\log p_t^i$, where the price was defined to be constant between consecutive trades (PTE prescription).
All series have been normalized to zero mean and unit variance. It has been assumed that measured prices are randomly sampled points of an underlying synchronous time series as described
above. The sampling  rates $\lambda_i$ were computed for each stock and the waiting time distributions have been taken to be exponential as a first order approximation.

\begin{figure*}
\begin{center}
\includegraphics[width=\linewidth]{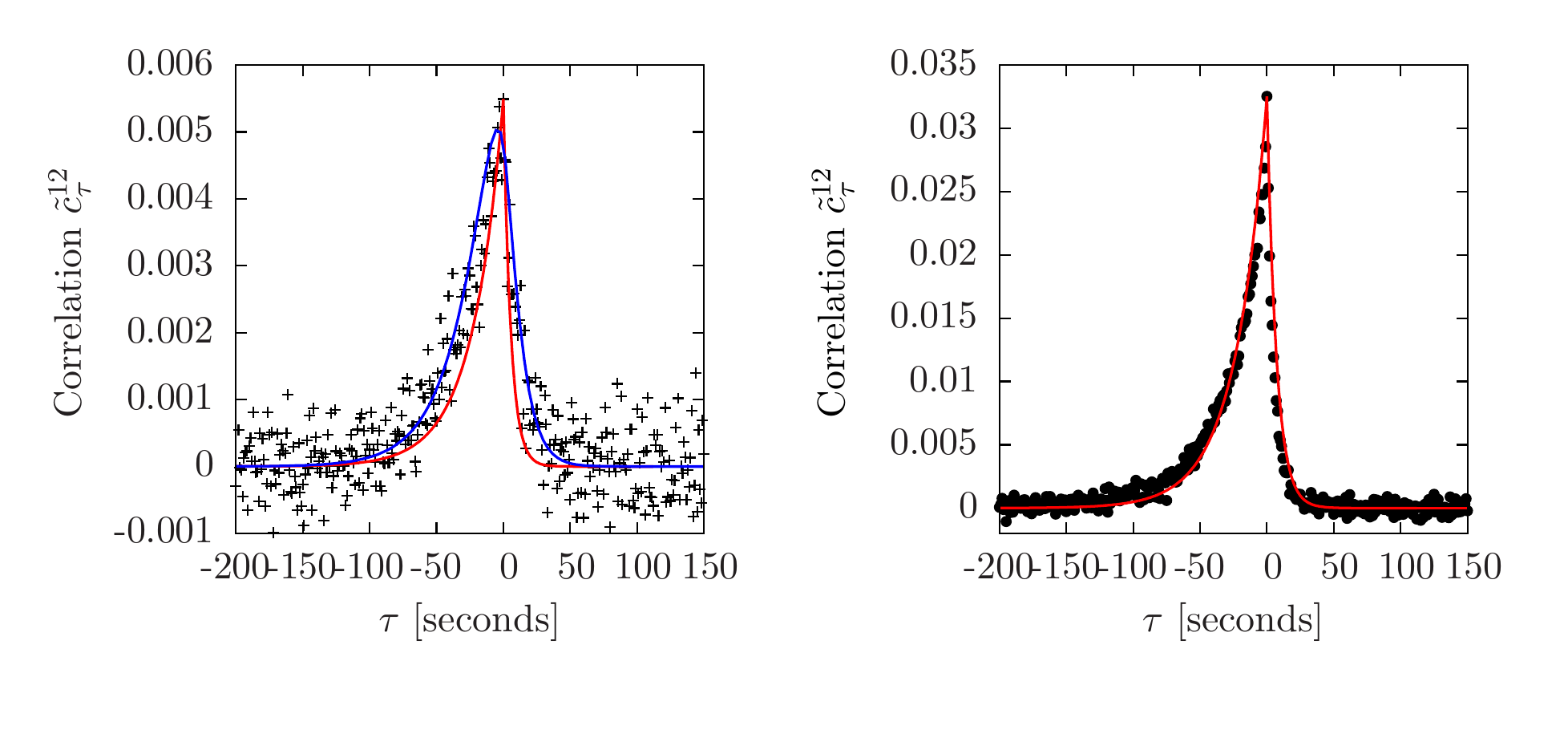} 
\caption{Left: The raw, infinitesimal cross-correlation coefficient $\tilde c^{\, GE/K}_\tau$ is shown for the pair of assets GE and K as a function of the lag $\tau$ (black points); the asymmetry of this function can be 
explained assuming genuine correlations of the simple form $ c^{GE/K}_{\tau} = c \, \delta_\tau$ and convoluting the effect of the sampling (red line); the best fit of the form of
equation (\ref{CorrectCrossCorr}) 
is also
shown (blue line). Right: Infinitesimal cross-correlation coefficient $\tilde c^{12}_\tau$ for two asynchronously sampled processes (black points); the evolution of the underlying time series with constant correlation was simulated. Sampling times were taken to coincide with those of the stocks GE and K in the data set. The
red line shows the theoretical correlation curve obtained for the same underlying process with an exponential waiting time distribution matching the measured sampling rates.}
\label{FigGEK}     
\end{center}
\end{figure*}

Cross-correlation coefficients have been systematically calculated; 
as expected raw cross-correlation coefficients $\tilde c^{ij}_\tau$ show a narrow peak near $\tau=0$ corresponding to the market mode (figure \ref{FigGEK}), justifying a fit with functions of the form:
\begin{equation}
\label{RawCrossCorr}
\tilde c^{ij}_\tau = c_{sync} \, e^{- |\tau - \tau_{sync}|/\xi_{sync}}
\end{equation}
The influence of the asynchronous sampling on these inferred parameters is indeed relevant, as the typical sampling times $\lambda^{-1}$ are of the same order of $\xi_{sync}$; the
simplest way to take into account its effect is to fit using functions of the form:
\begin{equation}
\tilde c^{ij}_\tau = c_{asyn} \, e^{- |\tau - \tau_{asyn}|/\xi_{asyn}} * K^{ij}_\tau
\label{CorrectCrossCorr}
\end{equation}
where $K^{ij}_\tau$ is the kernel appearing in (\ref{Convolution}), which depends on the estimated sampling frequencies $\lambda_i$ and $\lambda_j$, and $*$ denotes convolution. 

\begin{figure}
\begin{center}
  \includegraphics[width=3in]{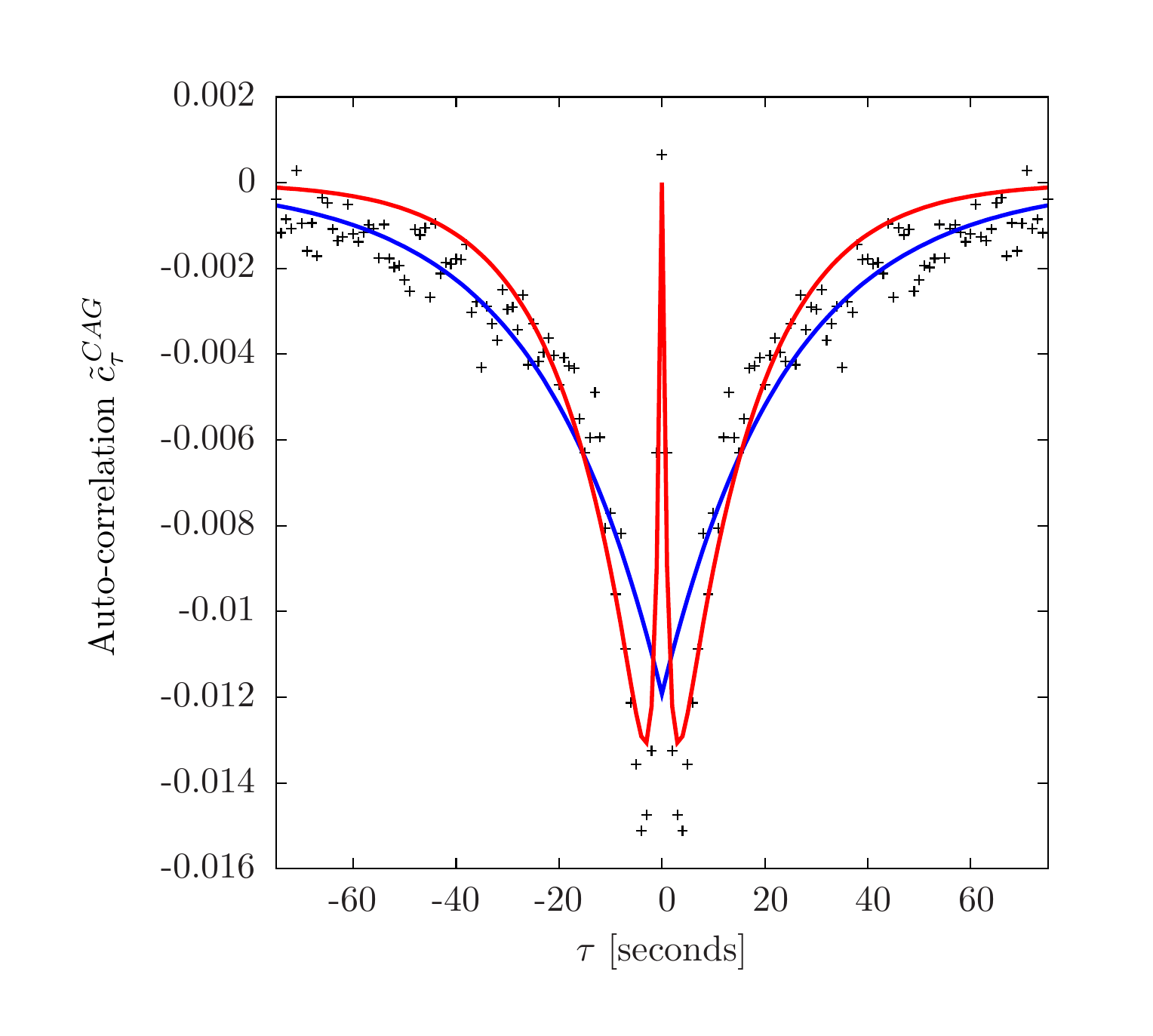}
\caption{A typical infinitesimal auto-correlation coefficient (in this case $\tilde c^{\, CAG}_\tau$) is plotted (black points). Its best fit of the form (\ref{RawAutoCorr}) is plotted in blue,
while the best fit of the form (\ref{CorrectAutoCorr}) is represented in red. Notice that even if the empirical function we plotted is negative and has a bimodal shape,
a positive diverging contribution in $\tau = 0$ should also be taken into account.}
\label{Fig5}       
\end{center}
\end{figure}

Auto-correlations have also been computed for all the stocks, and both their qualitative and quantitative behavior turn out to be very different from the case of cross-correlations.
In particular one can see that all auto-correlations are positively divergent in the origin, but assume finite values for lags different than 0, as shown in figure \ref{Fig5}.
Then the simplest fit that can be performed is the one with a function of the kind:
\begin{equation}
\label{RawAutoCorr}
\tilde c^{ii}_\tau = a_{sync} \, \delta_\tau - b_{sync} \left( \frac{e^{-|\tau|/\xi_{sync}}}{2 \, \xi_{sync}} \right)
\end{equation}
which is the superposition of a purely Brownian part with a fast decaying part. As in the case of cross-correlations, we can also fit those functions using their asynchronous counterpart:
\begin{eqnarray}
\label{CorrectAutoCorr}
\tilde c^{ii}_\tau &=& \left(a_{asyn} - \frac{b_{asyn}}{1+\lambda_i \xi_{asyn}} \right) \delta_\tau \\
&-& b_{asyn} \left[ \frac{\xi_{asyn} \lambda_i^2}{2[(\lambda_i \xi_{asyn})^2 - 1]} \left( e^{-|\tau| / \xi_{asyn}} - e^{-\lambda_i |\tau|} \right) \right] \nonumber  \;,
\end{eqnarray}
as suggested by the examples discussed in the previous sections.
The results of the fit of auto- and cross-correlation coefficient  with the raw and corrected functions defined above are summarized in Table \ref{tab:1}, where three kinds of ensembles (AC, T and L)
were considered.

\begin{table}
\caption{Results for auto- and cross-correlation coefficients $\tilde c^{ij}_\tau$ fitted against the functions defined in section \ref{EmpirSection} for various ensembles. Ensemble AC, used to compute
auto-correlations, contains the 100 most traded assets of the NYSE, while ensembles T and L contain, respectively, the 10 more traded and the 10 less traded assets of the same market, and have been used
to compute cross-correlation functions. For each of the parameters we write the ensemble average and show in parenthesis its standard deviation.}
\label{tab:1}       
\begin{tabular}{lccccc}
\hline\noalign{\smallskip}
Ensemble & $\xi_{sync} $ & $\xi_{asyn} $& $\tau_{sync} $ & $\tau_{asyn} $ & $  \frac{\chi^2_{sync}}{ \chi^2_{asyn}} - 1$ \\
\noalign{\smallskip}\hline\noalign{\smallskip}
AC   & 21.8    & 1.27 & \quad - & - & 0.30 \\
         & (32.3) & (1.36) & \quad - & - & $(0.63)$\\
\noalign{\smallskip}\hline\noalign{\smallskip}
T-T & 12.93 & 7.69 & 0.30 & -0.27 &  -0.05\\
 & (1.56) & (2.07) & (1.55) & (1.98) & (0.14)\\
T-L & 21.42 & 9.36 & 8.62 & 2.10 &  0.69 \\
 & (3.99) & (4.45) & (4.62) & (4.04) & (0.36)\\
L-L & 28.36 & 10.85 & -0.73 & -1.66 &  0.005\\
& (4.60) & (6.13) & (4.14) & (4.98) & $(0.08)$\\
\noalign{\smallskip}\hline
\end{tabular}
\end{table}

\begin{figure*}
\begin{center}
  \includegraphics[width=5in]{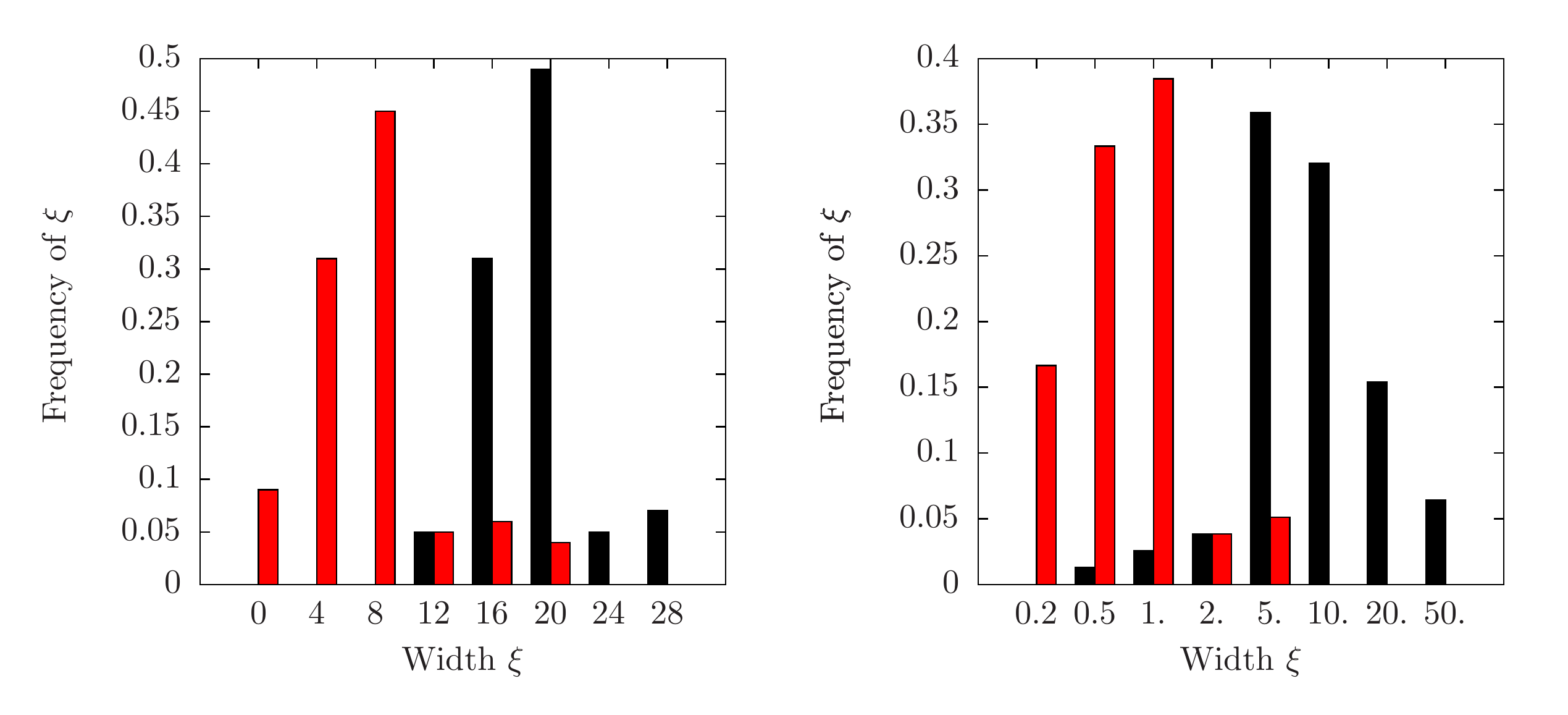}
\caption{Histogram of the fitted values of $\xi_{sync}$ (black bars) and $\xi_{asyn}$ (red bars), both for cross-correlations (left) and auto-correlations (right).
In the case of cross-correlations we considered a sample consisting of the 10 more active assets and the 10 less active ones, while for the auto-correlation we
considered the 100 most active assets. Notice that while the left plot is in linear-linear scale, the right one is in log-linear scale: for auto-correlations
most of the width is induced by the sampling, while for the cross-correlations the asynchrony seems to play a less significant r\^ole.}
\label{Fig6}     
\end{center}
\end{figure*}

First we discuss the results for the ensemble AC, which contains the 100 most traded assets of NYSE, and has been used to compute the infinitesimal auto-correlation coefficients.
The raw functions have a raw width $\xi_{sync}$ broadly distributed around a mean value of 20 s, as shown in Table \ref{tab:1}, and are typically characterized by a bimodal shape
(79\% of the empirical functions are compatible with zero for $\tau = 0$) which the raw model cannot account for. The inclusion of the sampling effect in the fitting functions improves
the descriptive power of the model just slightly: on average the chi-square is reduced of about 30\%, but fluctuations in the ensemble are strong. Indeed, the asynchrony explains naturally the
bimodal shape of the auto-correlations, and shifts the width of the corrected function $\xi_{asyn}$ to a small interval centered around a value of 1 s (figure \ref{Fig6}), providing thus a mechanism to
explain most of the signal width.
A similar result holds for the ratios $a_{sync}/b_{sync}$ and $a_{asyn}/b_{asyn}$: while the former follows a broad distribution, the latter is sharply peaked around a mean value of $\approx 1.5$.
These results do not qualitatively change if one takes as synchronous fitting function the superposition of a delta function with two exponentials.

The other ensembles which have been considered are T and L, containing respectively the 10 most and less traded assets of the AC ensemble; they have been used to calculate the infinitesimal
cross-correlation coefficient for all the pairs of the form T-L, T-T and L-L.
The inferred widths $\xi_{sync}$ are generally spread on a window of 30 s, ranging from 10 seconds (T-T ensemble) to 40 seconds (L-L), while the values of $\tau_{sync}$ often exceed 10 s in the T-L case,
indicating that a lack of symmetry is present in this ensemble; the direction of the asymmetry reveals an influence of the most traded stocks towards the less traded ones.
For the T-L ensemble, the asynchronous model turns out to provide a better description of the data (see e.g. figure \ref{FigGEK}), as residuals are significantly reduced, and most of the asymmetry is accounted
for in the kernel. Additionally, asynchronous sampling explains much of the observed width of correlations functions, as shown in figure \ref{Fig6}. Still, compared to auto-correlations the histogram of estimated
widths of cross-correlations  is centered at values significantly different from zero, of the order of 10 seconds.
In the T-T and L-L cases the descriptive power of the two models is almost identical (when sampling rates are similar, it becomes harder to statistically discriminate functions
(\ref{RawCrossCorr}) and (\ref{CorrectCrossCorr})). Again, even if a part of the width $\xi_{sync}$ is explained by sampling, the value of $\xi_{asyn}$ is significantly different from zero, meaning that
other mechanisms contribute to the formation of Epps effect. Interestingly, while the raw width varies significantly within the ensembles T-L, T-T and  L-L, the corrected width $\xi_{asyn}$ is compatible for all of
them and of the order of 10 seconds.

\begin{figure}
\begin{center}
  \includegraphics[width=3in]{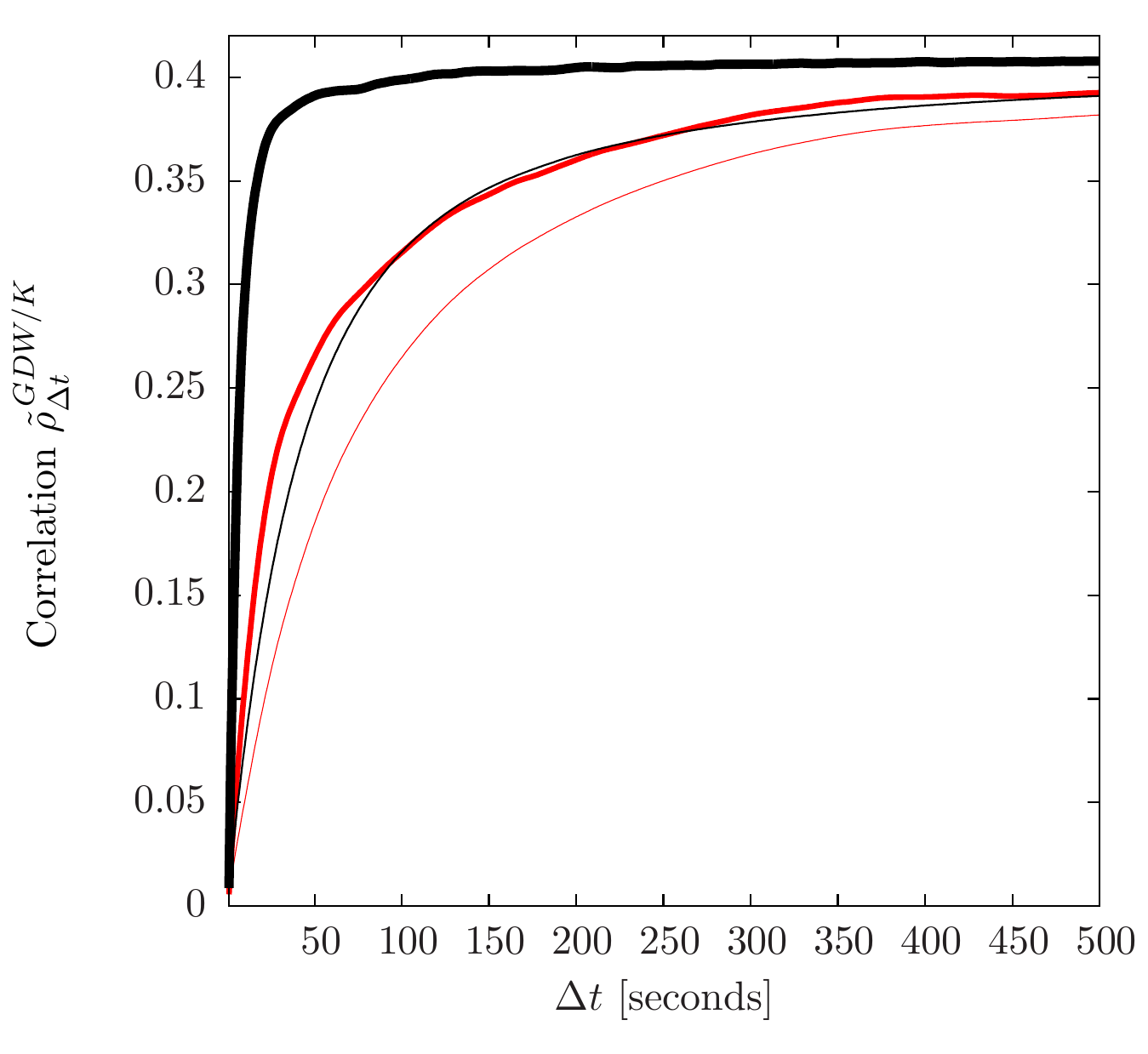}
\caption{The raw equal time correlation coefficient $\tilde \rho^{\, GDW/K}_{\Delta t}$ (narrow red line) is shown for the pair GDW and K, together with the same curve obtained with filtered
data (thick red line).
The black line corresponds to the correlation coefficient for a simulated process with the same asymptotic value of $\tilde \rho^{12}_{\Delta t}$, sampling rates and statistics of the other curves,
whose cross and auto-correlations are $\propto \delta_\tau$; the dashed line is the filtered version of the same curve.}
\label{Fig7}
\end{center}
\end{figure}

\begin{figure}
\begin{center}
  \includegraphics[width=3in]{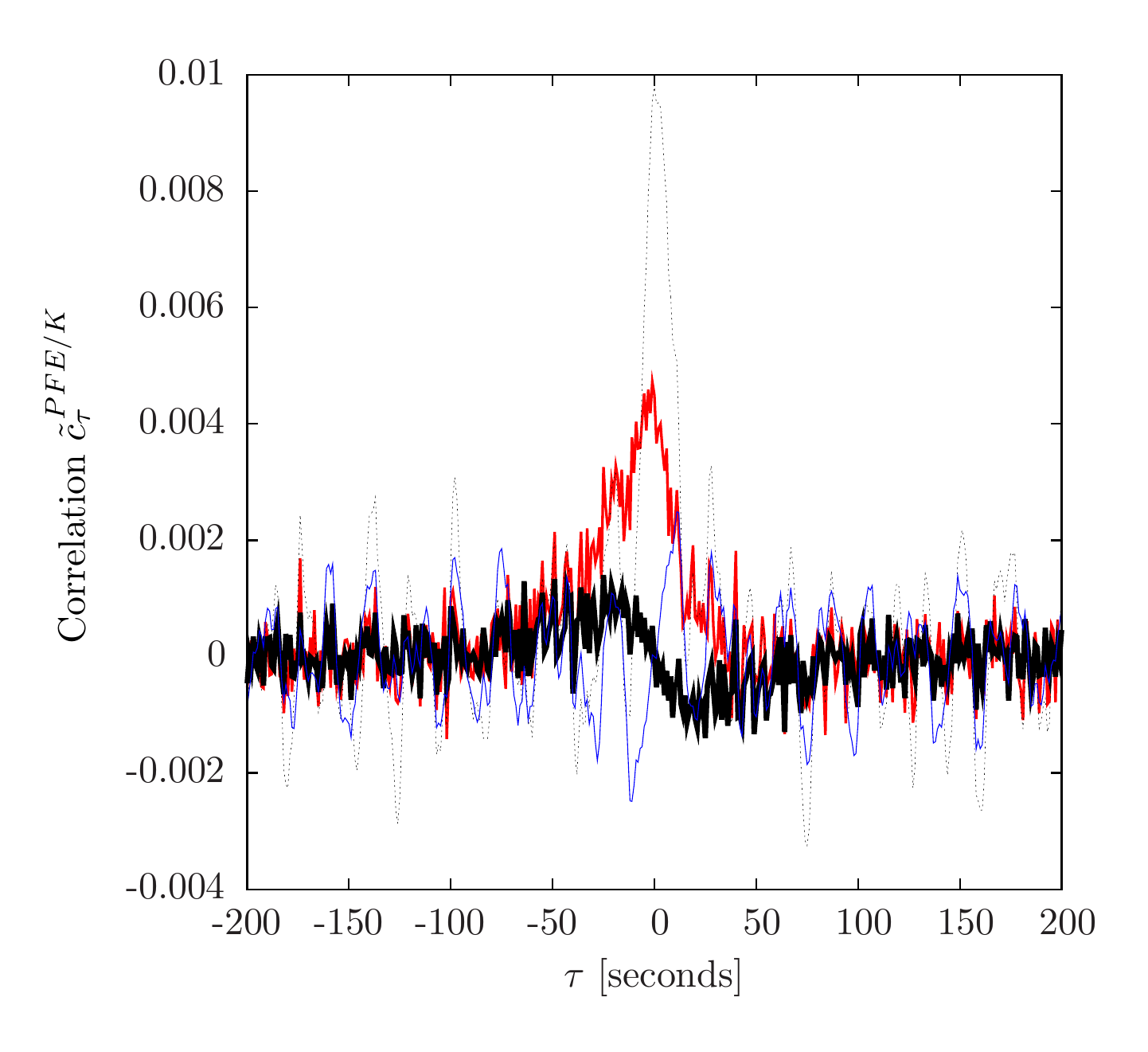}
\caption{Raw and filtered infinitesimal cross-correlation coefficients $\tilde c^{\, PFE/K}_\tau$ for stocks PFE and K calculated in year 2003 (respectively, red and dotted line), together with their 
asymmetrical parts $(\tilde c^{\, PFE/K}_\tau - \tilde c^{\, PFE/K}_{-\tau})/2$ (black and blue line).}
\label{Fig8}
\end{center}
\end{figure}

In order to compensate for the effect of the sampling it is also possible to filter the raw signal using the procedure described above; this allows us to evaluate the impact of the asynchrony on the
measured correlations as a  function of the scale $\Delta t$. Figure \ref{Fig7} shows the saturation curves of the correlation $\rho^{ij}_{\Delta t}$ for a pair of assets using both raw and filtered data
and compares them with the ones obtained for a pair of simulated Brownian motions with the same asymptotic value of correlation. Results obtained for simulated data set the maximum efficiency
of the filter, which is fixed by the length of the time series; empirical data show that the reconstructed curve is well below such bound, indicating that other effects do contribute to the formation of the
Epps. These features include by micro-structural effects, such as finite tick-size \cite{Munnix2010a} \cite{Munnix2010b}, and possibly an intrinsic time
scale related to human reaction \cite{Toth2009a} .
The same features are detected in figure \ref{Fig8}, where the infinitesimal, raw cross-correlations $\tilde c^{ij}_\tau$ are compared to the filtered ones; the presence of a residual Epps effect is
indicated here by the finite width of the filtered curve. Additionally, most of the asymmetry contained in the raw curve can be successfully removed, as most of the lag is induced as an effect
of the different sampling rates.

Within this approach it is also necessary to estimate from empirical data the nature of the waiting time distribution, as it usually deviates from the exponential one which is assumed.
The effect of the deviations must then be evaluated to ensure the consistency of the procedure previously described. We analyzed this issue by simulating a set of synchronous time series of known spectrum, and 
sampling them using points extracted from real data; then spectra for those series were systematically checked against the analytical predictions obtained for an exponential waiting time distribution. On the right side of figure \ref{FigGEK} we 
compare the effect of an exponential sampling with the real one, finding that no significant difference is induced by fluctuations of $\lambda_i$.

\section{Conclusions}
\label{FinalSection}
In this paper we investigated the time-dependence of financial correlations and their decay at very high frequency (Epps effect),
showing that some simple models of stochastic process are able to describe this features.
We found  that in case of exponentially sampled data the impact of asynchrony on correlations can be analytically
controlled, and its contribution can be exactly evaluated.
We also find that within this framework one can successfully describe some features of the empirical correlations observed
in the NYSE market, namely the heterogeneity of the price change predictability  and the presence of a causal
structure in the cross-correlations.
The first feature is detected as a broad distribution of widths both in the auto- and cross-correlation functions of the assets, and
can be explained by taking into account the effect of the sampling. The second one is quantified by the lag of cross-correlation
functions, and again can be almost completely justified by including the sampling effect.
Finally, we find that a significant fraction of the Epps effect cannot be explained as just due to the effect of asynchrony,
indicating that other kind of effects, conjectured in \cite{Toth2009a} to be related to time scales of human reaction,
contribute to the observed dynamics of correlations.

\appendix

\section{Effect of an exponential random sampling}
\label{AppendixB}
We now turn to prove the properties which allow us to analytically account for the effect of the random sampling.\\
To prove property P1, we consider a multivariate synchronous process $X^i_{\Delta t}$, and let the asynchronous sampling be induced by a waiting time distribution
$p_i(t) = \lambda_i e^{-\lambda_i t}$ as described in  section $\ref{AsyncSection}$. We want to show that for such a process, the covariance can be computed using the substitution:
\[
\tilde S^{ij}_\omega = S^{ij}_\omega  \frac{\lambda_i \lambda_j}{(\lambda_i + i \omega)(\lambda_j - i \omega)}
\]
where $S^{ij}_\omega$ is the spectrum of the synchronous process. This can be seen by directly computing the covariance, which is by definition:
\begin{eqnarray*}
\tilde C^{ij}_{\Delta t} &=& E  \left[ \int_{t^i_1}^{t^i_2} \int_{t^{\prime j}_1}^{{t^{\prime j}_2}} \langle dX^i_t dX^j_{t^\prime} \rangle \right] \\
&=&    \lambda_i^2  \lambda_j^2 \int_{-\infty}^0 dt^i_1 dt^{\prime j}_1 \int_0^{\Delta t} dt^i_2  dt^{\prime j}_2 \nonumber \left( \int_{t^i_1}^{t^i_2} \int_{t^{\prime j}_1}^{{t^{\prime j}_2}} \langle dX^i_t dX^j_{t^\prime} \rangle \right)
 \; e^{\lambda_i (- \Delta t + t^i_1  + t^i_2 )} \; e^{\lambda_j (- \Delta t + t^{\prime j}_1  + t^{\prime j}_2 )} \;,
\end{eqnarray*}
where we have used the symmetry with respect to time inversion of the exponential waiting time distribution. The expression in parenthesis can be written in Fourier space as:
\[
 \int_{t^i_1}^{t^i_2} \int_{t^{\prime j}_1}^{{t^{\prime j}_2}}  \langle dX^i_t dX^j_{t^\prime} \rangle  =  \frac{1}{2 \pi} \int_{t^i_1}^{t^i_2} dt \int_{t^{\prime j}_1}^{{t^{\prime j}_2}} dt^\prime \int d\omega \, S^{ij}_\omega e^{- i \omega (t - t^\prime)}
\]
And the two time integrals can be performed, leading to:
\[
\frac{1}{2 \pi}  \int d\omega \frac{S^{ij}_\omega}{\omega^2} \left( e^{-i \omega t^i_2} - e^{-i \omega t^i_1} \right) \left( e^{i \omega t^{\prime j}_2} - e^{i \omega t^{\prime j}_1} \right)
\]
Now one can integrate over the waiting time measure, getting as a final expression:
\[
\tilde C^{ij}_{\Delta t} = \frac{1}{2 \pi}  \int d\omega \frac{S^{ij}_\omega}{\omega^2} \frac{\lambda_i \lambda_j}{(\lambda_i + i \omega)(\lambda_j - i \omega)}
  \left( e^{-i \omega \Delta t} - 1 \right) \left( e^{i \omega \Delta t} - 1 \right) \nonumber
\]
which is identical to equation (\ref{Covariance}) obtained in the synchronous case, except for the substitution \\ $S^{ij}_{\omega} \rightarrow  S^{ij}_{\omega}  \frac{\lambda_i \lambda_j}{(\lambda_i + i \omega)(\lambda_j - i \omega)}$. In this last step the presence of an exponential sampling is crucial to obtain a convolution as the result of the computation, as the dependence of the above integrand from $\Delta t$ requires a cancellation; in
particular one can see that the exponential waiting time distribution is the only one producing a convolution as the result of this last integration. It is also important to remark that independence between the sampling 
process and the underlying time series has been implicitly assumed in all our construction.

Also P2 can be proved by directly calculating the variance. In particular, if given a synchronous process  of spectrum $S_\omega$ one builds an asynchronous process of sampling rate $\lambda$, its
variance is given by:
\begin{eqnarray}
\tilde C_{\Delta t}&=& E\left[ \frac{1}{2 \pi} \int_{-\infty}^{+\infty} d\omega\, \frac{S_\omega}{\omega^2} \left( 2 - e^{-i \omega (t_2-t_1)} - e^{i \omega (t_2 - t_1)}\right) \right] \nonumber \\
&=&  \frac{1}{2 \pi} \int_{-\infty}^{+\infty} d\omega \, \frac{S_\omega}{\omega^2} \bigg\{ \lambda^2 \int_0^\infty \int_0^{\Delta t} d\tau_1 d\tau_2 \, e^{-\lambda \tau_1} e^{-\lambda \tau_2} \nonumber  \\
&\times&\left( 
2 - e^{-i \omega (\Delta t - \tau_2 + \tau_1)} - e^{i \omega (\Delta t - \tau_2 + \tau_1)} \right) \bigg\}  \nonumber 
\end{eqnarray}
which results:
\begin{eqnarray}
\tilde C_{\Delta t} &=& \frac{1}{2 \pi} \int_{-\infty}^{+\infty} d \omega \, \frac{S_\omega}{\omega^2} \left( e^{- i \omega \Delta t} -1 \right) \left( e^{ i \omega \Delta t} -1 \right) + \nonumber   \\
&+& \frac{2}{\lambda^2} \left[  \frac{1}{2 \pi} \int_{-\infty}^{+\infty} d \omega \, \frac{S_\omega}{1+ \omega^2/\lambda^2}  \left( e^{-i \omega \Delta t} - e^{-\lambda \Delta t} \right) \right]  \nonumber 
\end{eqnarray}

Finally, if variances in the synchronous case are linear (i.e. $\langle (X_{\Delta t})^2 \rangle= \sigma^2 \Delta t$) or equivalently if $S_\omega$ is constant, then in the asynchronous one they
are not modified, as one can see computing the correcting term in equation (\ref{VarianceSubstitution}), which in this case is vanishing.

\section{Calculation of variance and covariance}
\label{AppendixA}
Given a synchronous process $X^i_t$, we are interested in calculating the quantities $C^{ij}_{\Delta t}$ and $\tilde C^{ij}_{\Delta t}$ defined as in equation (\ref{Covariance}) and
(\ref{AsyncCovariance}) in some representative cases. Indeed we will write the expression for the asynchronous covariance only, considering exponential waiting time processes
of rates $\lambda_i$, as the corresponding expressions for the synchronous case can be obtained taking the limit $\lambda_i \rightarrow \infty$ in the resulting formulas.
Let us consider for the synchronous process a spectrum of the kind:
\[
S^{ij}_\omega = \frac{e^{i \omega \tau}}{1+\omega^2 \xi^2}
\]
where we assume $\tau > 0$ and $\xi>0$, consistently with the assumption of correlations of the kind $c^{ij}_{t-t^\prime} = \frac{1}{2\xi} e^{-|t-t^\prime-\tau|/\xi}$,
in which a lag and an exponential decay are superimposed. Then one can calculate using equation (\ref{Covariance}) and substitution (\ref{SpectrumSubstitution}):
\[
\tilde C^{ij}_{\Delta t} = \frac{1}{2\pi} \int_{-\infty}^{+\infty} \frac{d\omega }{\omega^2} \left( e^{- i \omega \Delta t} -1 \right) \left( e^{ i \omega \Delta t} -1 \right)
 \left[ \frac{e^{i \omega \tau}}{(1+\omega^2 \xi^2)(1+i\omega/\lambda_i)(1-i\omega/\lambda_j)} \right]
\]
Above integral can be solved by integration on the complex plane after choosing an appropriate contour. In particular the integral can be written as:
\[
\tilde C^{ij}_{\Delta t} = \frac{1}{2\pi} \int_{-\infty}^{+\infty} d\omega \; (A^{ij}_\omega + B^{ij}_\omega)
\]
with:
\begin{eqnarray*}
A^{ij}_\omega &=&\phantom{-} \left( \frac{2 - e^{i \omega \Delta t} }{\omega^2} \right)  S^{ij}_\omega \\
B^{ij}_\omega &=&-  \left(  \frac{ e^{-i \omega \Delta t} }{\omega^2} \right)  S^{ij}_\omega
\end{eqnarray*}
Then the full integral can be splitted in two (diverging) parts, whose value can be calculated by residues. In particular, while the integral of $A^{ij}_\omega$ can be always found by choosing a semicircular
 contour closed on the upper imaginary plane, to integrate $B^{ij}_\omega$ it is necessary to close the contour according to the sign of $\Delta t - \tau$. Then the result splits into:
 \begin{eqnarray}
 \tilde C^{ij}_{\Delta t} &=&  {\rm Res}_{ i /\xi } A_\omega + {\rm Res}_{i \lambda_i} A_\omega - {\rm Res}_{ -i /\xi } B_\omega \nonumber \\
 &-& {\rm Res}_{ -i \lambda_j } B_\omega + {\rm Res}_{ 0 } (A_\omega - B_\omega)/2 \nonumber
 \end{eqnarray}
 for $\Delta t > \tau$, and:
 \begin{eqnarray}
 \tilde C^{ij}_{\Delta t} &=&  {\rm Res}_{ i /\xi } A_\omega + {\rm Res}_{i \lambda_i} A_\omega + {\rm Res}_{ i /\xi } B_\omega  \nonumber \\
 &+& {\rm Res}_{ i \lambda_i } B_\omega + {\rm Res}_{ 0 } (A_\omega +  B_\omega)/2   \nonumber
 \end{eqnarray}
for $\Delta t < \tau$, where ${\rm Res}_{ z_0} f_z$ denotes the residue of $f_z$ in $z_0$. For  $\Delta t > \tau$ this reads:
\begin{eqnarray}
\tilde C^{ij}_{\Delta t} &=& \Delta t - \tau + \lambda_i^{-1} - \lambda_j^{-1} \nonumber
+ \lambda_i \lambda_j \xi^3 \left( \frac{e^{- (\Delta t - \tau)/\xi }}{2 u_i v_j } - \frac{e^{- \tau/\xi }}{v_i u_j} + \frac{e^{- (\Delta t + \tau)/\xi }}{2 v_i u_j} \right) + \nonumber  \\
&+& \left[ \frac{\lambda_j e^{- \lambda_i \tau}}{\lambda_i (\lambda_i + \lambda_j) u_i v_i} \right]  (2 - e^{- \lambda_i \Delta t}) \nonumber 
- \left[ \frac{\lambda_i e^{- \lambda_j (\Delta t - \tau)}}{\lambda_j (\lambda_i + \lambda_j) u_j v_j} \right]  \nonumber 
\end{eqnarray}
while for $\Delta t < \tau$ it is:
\[
\tilde C^{ij}_{\Delta t} = \frac{\lambda_i \lambda_j \xi^3 e^{-\tau/ \xi}}{v_i u_j} \left( \cosh ( \Delta t/\xi) - 1 \right)
+ \left[ \frac{2 \lambda_j e^{- \lambda_i \tau}}{\lambda_i (\lambda_i + \lambda_j) u_i v_i} \right]  (1 - \cosh (\lambda_i \Delta t)) \;,
\]
where $u_i = 1 + \lambda_i \xi$ and $v_i = -1 + \lambda_i \xi$.
Formulas given in the examples of section \ref{SyncSection} and  \ref{AsyncSection} can be recovered from this expression by taking the appropriate limits. \\ \\

\section{Finite size effects}
\label{AppendixC}
The construction described in sections  \ref{SyncSection} and \ref{AsyncSection} can be generalized to the case of finite size time series in discrete time ($t=1,\dots,T$): after having defined the discrete Fourier transform of the series $dX^i_t$ as:
\begin{eqnarray*}
dX_n ^i &=& \sum_{t=0}^{T-1} \, dX^i_t \, e^{2 \pi i n t / T}  \\
dX_t^i &=& \frac{1}{T} \sum_{n=0}^{T-1} \, dX^i_n \, e^{- 2 \pi i n t / T}
\end{eqnarray*}
and the spectrum as:
\[
S^{ij}_n = \frac{\langle dX^i_n dX^j_n \rangle}{T} \,,
\]
it is possible to consider an asynchronous sampling defined through a rate $\Lambda_i$, so that the probability of sampling the time series at time $t$ is uniform and equal to $1 - e^{-\Lambda_i}$. In this case the sampling induces an analogous effect, and substitution (\ref{SpectrumSubstitution}) becomes:
\[
\tilde S^{ij}_n = S^{ij}_n  \left( \frac{1 - e^{-\Lambda_i}}{1-e^{-\Lambda_i + 2 \pi i n / T}} \right) \left(\frac{1 - e^{-\Lambda_j}}{1-e^{-\Lambda_j - 2 \pi i n / T}} \right)
\]
The filtration procedure described in section \ref{FilterSection} can still be applied in this case, where it is affected by a finite error: correlations in real time
have a minimum resolution which scales as $T^{-1/2}$, as noise effects on the measured spectrum fix the accuracy of the reconstructed signal.

%
%
%

%
%

\end{document}